\documentclass[final,5p,times,twocolumn,authoryear]{elsarticle}

\usepackage{amssymb}

\usepackage[table,xcdraw]{xcolor}
\usepackage[normalem]{ulem}
\usepackage{array}
\newcolumntype{x}[1]{>{\centering\arraybackslash\hspace{0pt}}p{#1}}

\definecolor{ao}{rgb}{0.0, 0.5, 0.0}

\usepackage{enumitem}
\setlist[description]{leftmargin=1em}

\usepackage[T1]{fontenc}

\PassOptionsToPackage{hyphens}{url}
\usepackage[hidelinks]{hyperref}
\hypersetup{
    colorlinks,
    linkcolor=blue,
    citecolor=blue,
    urlcolor=blue
}
\expandafter\def\expandafter\UrlBreaks\expandafter{\UrlBreaks
    \do\a\do\b\do\c\do\d\do\e\do\f\do\g\do\h\do\i\do\j%
    \do\k\do\l\do\m\do\n\do\o\do\p\do\q\do\r\do\s\do\t%
    \do\u\do\v\do\w\do\x\do\y\do\z\do\A\do\B\do\C\do\D%
    \do\E\do\F\do\G\do\H\do\I\do\J\do\K\do\L\do\M\do\N%
    \do\O\do\P\do\Q\do\R\do\S\do\T\do\U\do\V\do\W\do\X%
    \do\Y\do\Z\do\*\do\-\do\~\do\'\do\"\do\-}%

\newcommand{\urlDate}{last accessed 2021-02-22}
\newcommand{\furl}[1]{\footnote{\url{#1} (\urlDate).}}

\journal{DFRWS EU 2026}

\begin{document}

\begin{frontmatter}



\title{Effects of parental controls in the context of Digital Forensics}

\author[label1]{Selina Märchy}
\author[label1]{Mauro Vignati}
\author[label2]{Frank Breitinger}

\affiliation[label1]{organization={University of Applied Sciences Bern},
            country={Switzerland}}

\affiliation[label2]{organization={University of Augsburg},
           country={Germany}}

\begin{abstract}
Parental control systems are designed to protect minors online, but can inadvertently obstruct digital forensic investigations. When enabled, these systems restrict administrative privileges, disable debugging options, and alter data accessibility, complicating evidence acquisition and analysis. This study empirically examines the impact of Microsoft, Google, and Apple parental controls on forensic processes across fifteen Windows, Android, and iOS devices.
Through controlled experiments, we evaluate their impact on evidence accessibility and identify forensically sound methods to overcome these limitations. The findings provide practical guidance for investigators and contribute to improving forensic readiness in environments governed by parental control systems.
\end{abstract}




\end{frontmatter}



\section{Introduction}
\label{introduction}
To protect minors from online risks and inappropriate content, parental control systems offered by prominent technology providers such as Microsoft, Google, and Apple have become widely adopted \citep{wang2021protection}. These systems allow parents to monitor, restrict, or filter their children's digital activities, yet their effects extend beyond everyday use. When a minor becomes involved in a criminal investigation, either as a suspect, victim, or witness, these controls can complicate digital forensic procedures. Parental control configurations often limit administrative privileges, disable debugging modes, or restrict data access, much like enterprise Mobile Device Management (MDM) solutions \citep{barmpatsalou2018current}. Consequently, forensic practitioners may face difficulties acquiring or analyzing evidence in a forensically sound manner. 

\paragraph{Related work} While there has been research on parental control systems, we could not find any directly related work. 
Prior works discussed the security and privacy risks of these systems \citep{ali2020betrayed,feal2020angel}, requirements for tools \citep{assis2024my}, how to develop more inclusive and effective online safety technologies \citep{iftikhar2021designing}, or how it affects the well-being of parents \citep{bertrandias2023using}. In addition, researchers focused generally on the impact of parental control of technology on the children \citep{richardson2021longitudinal}. Lastly, there have been attempts to create independent software solutions for parental control, such as Guardian Eye by \cite{joshi2024guardian}. 
Despite extensive research on parental control technologies and digital forensics \citep{luciano2018digital,breitinger2024dfrws} individually, little is known about how these systems impact during forensic acquisition and analysis.

\paragraph{Research gap} This study addresses that gap by examining how native parental control mechanisms from Microsoft, Google, and Apple affect digital forensic processes across multiple device types. Specifically, we explore the following key questions: 
\begin{description}
    \item[RQ1:] To what extent do parental controls hinder the work of digital forensics?
    \item[RQ2:] To what extent do parental controls affect the quality of forensic images?
    \item[RQ3:] How can parental controls be disabled or bypassed in the most forensically sound manner?
\end{description}

\paragraph{Placement} The focus of this study is on parental control features integrated into operating systems, specifically Microsoft Family Safety, Google Family Link, and Apple ID for Children. Systems designed for specific children's devices (e.g., Lexibook) or specialized Linux distributions for children are outside the scope of this work. Similarly, manually configured parental controls, such as those deactivated or set up without OS-level integration, are excluded. Dedicated third-party parental control applications, including Net Nanny and Aura, are also not considered, as this study concentrates solely on controls embedded within the operating system environment. The study was conducted in Switzerland, and results may differ in other countries due to variations in local regulations and implementation practices.

\paragraph{Contribution} Through experiments on fifteen devices across Windows (4), Android (7), and iOS (4) ecosystems, this research provides an empirical assessment of the forensic implications of parental control technologies. The study contributes (1) a comparative analysis of forensic accessibility under parental control configurations, (2) practical guidance for investigators confronting restricted devices, and (3) recommendations for improving forensic readiness in environments where parental controls are active.

\paragraph{Outlook} Sec.~\ref{sec:parental_controls} provides background on parental control technology. The core of this article is Sec.~\ref{sec:experiment_phones} and \ref{sec:experiment_windows}, which discuss the experimental design, methodology, and results of this study for phones and Windows, respectively. In Sec.~\ref{sec:cloudtakeouts}, we discuss cloud takeouts, i.e., demanding data from cloud providers, which complements our findings. The limitations of this study are explained in Sec.~\ref{sec:limitations}. Sec.~\ref{sec:Key_Insights} comes back to the research questions and summarizes the key takeaways. The last section concludes the paper.

\paragraph{Note on Confidentiality and Disclosure}
Parts of this work are subject to non-disclosure agreements, and the used devices were mostly personal. 
Consequently, some details (e.g., names of some tools) and data outputs (e.g., screenshots/photos of settings) are omitted.
Researchers or practitioners requiring further clarification may contact the authors directly; additional information may be provided upon reasonable request and in accordance with confidentiality obligations.

\section{Parental control technology}
\label{sec:parental_controls}
This section provides background on parental control systems relevant to this study, i.e., Google Family Link \citep{stoev2023online}, Apple ID for Children, and Microsoft Family Safety \citep{franchuk2022using}. These are parental control systems that allow parents to manage their children's digital activity, but each functions within its own ecosystem. 

Overall, each platform offers similar parental oversight but is optimized for its respective devices and services.

The general process is similar across the three tested providers. 
One must install an application on the parent’s device, and then create or link the child's account. After that, one sets supervision permissions, and the child's device is managed using the settings defined by the parent.
All parental control systems define minors as individuals who have not reached the age of 13. Upon reaching this age, the young adult is entitled to assume full control over their account.

\subsection{Google Family Link}
\label{sec:googlefamilylink}
    Google Family Link is an application designed to safeguard the online activities of minors. It is accessible through the Google or Apple App Store. The control features only work on devices running Android 5.0 or higher, Fitbit Ace LTE, and Chromebook \citep{googlellcfamilylink2025}. Upon reaching 13, the young adult is entitled to assume full control over their Google account. Children below this age are permitted only to have a Google account with parental oversight capabilities. 
    Family Link lets parents supervise Google Accounts on Android and Chromebooks, controlling app downloads, screen time, and device location, though supervision is limited on iOS and Windows. 

 \subsection{Apple-ID for Children}
 \label{sec:appleidforchildren}
    To establish an Apple Family Group, an iPhone, iPad, or Mac device is required \citep{appledesign2025}. According to \cite{applescreentime2025}, the birth date of a child (account) under the age of 13 cannot be modified or altered, and it must mature naturally. Additionally, \cite{applefamilienfreigabe2024} offers an additional mode for young adults between 13 and 17 years. These individuals can be included in the family group but are not subject to child control restrictions. 
    Apple's system enables parents to create child Apple IDs and manage screen time, app access, content restrictions, and location tracking across iPhone, iPad, and Mac devices. 

 \subsection{Microsoft Family Safety}
 \label{sec:microsoftfamilysafety}
    Microsoft Family Safety is compatible with Xbox, Windows 10, and Windows 11 platforms. 
    Families \citep{microsoftfamilygrouproles2025} are organized with parents and children, where parents can set screen time limits, content filters, and tracking location, while providing activity reports.
    Since the account is not exclusively limited to children, the determination of whether a person is considered a child depends on their age \citep{microsoftmanagingconsentnd}. 
    According to \cite{microsoftfamilysafety2025}, Microsoft Family Safety is also accessible on Android devices. For administrative purposes, there are dedicated applications for Windows, Android, and iOS devices. 
     

In summary, Apple and Microsoft can natively control their devices more comprehensively, while Family Link mainly controls the Google ecosystem.

\subsection{Creating Accounts} 
\label{sec:creatingaccounts}

Google, Apple, and Microsoft Accounts for parents and children were created using fictitious details. We opted out of two-factor authentication whenever possible. 
For Windows accounts, an additional email or phone number was required. 

Certain mobile phones and specific phone service providers implement more stringent restrictions on children's accounts compared to those imposed on regular (parent) accounts:
\begin{itemize}
    \item On all iPhones, the installation of Telegram and Discord was not permitted for child accounts. The option to download these applications was deactivated, accompanied by a notification indicating that, for the safety of children, downloads are not allowed. 
    \item On Samsung phones, certain proprietary applications such as Samsung Notes, Samsung Health, and Samsung Voice were not pre-installed by default on a child's account. It was possible to download these applications with parental consent. Additionally, utilizing Samsung Health or performing backups via Samsung Cloud was not feasible, as creating a Samsung account is restricted for children. 
\end{itemize}

Regarding the parental control settings, the configurations were maintained at their default levels. As such, features like screen time limits, location tracking, and others were enabled and retained in their standard state. The sole modification involved extending the screen time to facilitate the comprehensive setup of the device. 

\section{Experimental design and results for mobile devices}
\label{sec:experiment_phones}
The overall objective of this study is to examine data acquisition capabilities across various smartphones (including different brands, iOS, and Android devices) within a controlled environment. Each device was tested under distinct operational conditions, such as immediately after first boot, to evaluate the extent of accessible data. Additionally, the study aimed to determine potential differences between parent mode and child mode configurations. For this purpose, each device was first used in parent mode (parent image), then factory reset and reconfigured in child mode, i.e., with active parent controls (child image). The parental control was implemented in accordance with the vendor's recommended standards. All usage activities were carefully documented, and every device was operated in an equivalent manner to ensure consistency across tests. 

This section first highlights some specific aspects relevant to the experiments, such as Android Debug Bridge (Sec.~\ref{sec:ADB}), various device states (Sec.~\ref{sec:BFU_AFU}). Subsequently, the methodology is explained in Sec.~\ref{sec:metho_phones} followed by a summary of the main findings.




\subsection{Android Debug Bridge (ADB)}
\label{sec:ADB}
The Android Debug Bridge (ADB) is a crucial interface for both Android development and digital forensic investigations. It enables direct communication between a computer and an Android device, allowing investigators to access the file system, extract data, and execute commands for analysis \citep{opasiakmazurczyk2018}. However, ADB functionality depends on Developer Mode being enabled on the device. Without Developer Mode, the device cannot establish a trusted connection with the host computer, and ADB commands cannot be executed. Consequently, enabling Developer Mode is a prerequisite for any forensic acquisition that relies on ADB communication. As \cite{reiber2019} emphasized:
 
``ADB must be used for all communications to and from an Android device, whether it is a logical or particular non-invasive physical collection. The Android device that is to be collected must have ADB enabled, either manually by the examiner or programmatically by the software.''

Typically, this setting is configured via Settings - Phone Information - Build Number - Developer Mode - USB Debugging.
If neither the AFU nor BFU variant is selected or available, enabling USB Debugging is required to obtain an ADB backup, which may serve as an alternative to a forensic image. As noted by \cite{reiber2019}, if the device is locked and ADB is not enabled, a logical image cannot be acquired, necessitating alternative methods such as physical collection with JTAG or obtaining root access \citep{breeuwsmajtag2006, ochengli2017}. Most of these methods are highly invasive and beyond the scope of this article.

\subsection{After First Unlock and Before First Unlock}\label{sec:BFU_AFU}
Mobile devices using file‑based encryption (FBE) operate in two key states for forensics: Before First Unlock (BFU) and After First Unlock (AFU). BFU occurs after a restart without entering the PIN. Only the PIN can decrypt files; biometric unlocks and some features, such as Wi‑Fi connections, are disabled, or the camera is unavailable. Data access is limited, making BFU the least favorable state for forensic analysis. If a device is in BFU state, the PIN/Pattern is commonly brute-forced \citep{campbell2023, classen2024}. 

AFU begins after the first PIN entry, unlocking encryption keys and enabling broader access to data until the next power off. In this state, exploits may be used to access the system, making brute-force attacks unnecessary. Automatic device restarts (iOS 18+, Android 2025) can force devices into BFU, affecting data accessibility \citep{classen2024, heise2025}. 

From a forensic perspective, having the PIN to access the device is ideal as it allows acquiring the data using ADB. 
AFU is preferable if the PIN is unavailable, as exploits may be used to obtain access. However, not all devices and software versions come with exploits, and there is a risk that the phone data may be damaged. A peculiarity is that using an exploit may allow imaging the system without Developer Mode/ADB being active.
Lastly, BFU is the least desirable as it requires brute-forcing the PIN, which may require a long time \citep{campbell2023, classen2024}. 
Once the PIN is found, one may proceed with an exploit (AFU) or with similar procedures as in the presence of the PIN (usually, tools make this decision where an exploit seems more common).


\subsection{Methodology}\label{sec:metho_phones}
The aim is to measure the impact of the parental control technology on the acquisition as well as how it may impact the analysis. The following steps were performed:

\begin{enumerate}
    \item \textbf{Initial Configuration (Parent):} The tested mobile phone was configured according to our playbook summarized in Sec.~\ref{sec:mobiledevices_conf} and left in parent mode.
    
    \item \textbf{Acquisition (Parent):} We acquired the data from the phone, i.e., triggering the imaging process. We simulated two situations (S):
    \begin{enumerate}
        \item S1: Unknown PIN code, AFU (the tool uses a known exploit to access the phone; usually does not require Android Debug Bridge (ADB) to be active).
        \item S2: Known PIN code (this requires ADB to be active to copy the content).
    \end{enumerate}
    Note, in two instances, AFU (S1) did not work. In these cases, we rebooted the device (BFU) and utilized the Brute-force functionality of the imaging software to get access.
    
    \item \textbf{Reset:} We performed a factory reset of the phone.
    
    \item \textbf{Initial Configuration (Child):} Like in the initial configuration (Parent) phase, we reconfigured the phone, but this time in child mode, i.e., the account changed, but the user interactions with the device remained the same (besides the exception discussed previously, such as installing Discord)
    
    \item \textbf{Acquisition (Child):} Similar to step 2, we copied the data from the phone under the identical conditions, i.e., S1 and S2. This allows us to compare the acquisition procedure. 

    \item \textbf{Data comparison:} An additional objective of this study was to determine whether parental control mechanisms affect the forensic analysis process and whether applications or system settings are accurately captured during imaging. Consequently, we manually compared the images that were acquired, i.e., the image parent vs.~image child, to see if there were any differences in data. 
\end{enumerate}

\subsection{Devices and their configuration} 
\label{sec:mobiledevices_conf}

Each device was configured with a standardized setup to ensure consistency across tests (details are provided in the Appendix\footnote{The configuration protocols (phones and Windows (see Sec.~\ref{sec:win_devices_conf}) are currently kept separate from this article and are attached for convenience; in case of acceptance, it may be integrated if the page limit allows it; otherwise, it will be hosted separately.}). Various types of data were loaded onto the devices, including pictures, videos, audio files, and documents (via USB for Android devices and iCloud for iPhones), as well as through email. Devices also generated their own media content using built-in cameras and microphones. Browser activity was equivalent by accessing the same websites, and chat applications (WhatsApp, Telegram, Signal) were installed and populated with messages, calls, and media, while WeChat was uninstalled. Social media apps (YouTube, Discord, TikTok, Snapchat, Instagram) were also installed, with content created in each. Additional data included address book entries, emails, calendar events, notes, health app data, maps, and phone calls. Some files were deliberately deleted to simulate typical usage. Geolocation permissions were granted, language was set to English, and the geographical location was Switzerland.
The following devices were used:

\begin{itemize}
    \item Apple iPhone 12 (A2172),  OS-Version: 18.3.2 (parent), 18.4.1 (child) 
    \item Apple iPhone 12 (A2172), OS-Version: 18.3.2 (parent), 18.4.1 (child) 
    \item Apple iPhone XR (A2105), OS-Version: 18.3.2 
    \item Apple iPhone 11 (A2221), OS-Version: 15.3.1 
    \item Samsung Galaxy A6 (SM-A600FN), OS-Version: 10
    \item Samsung Galaxy A51 (SM-A515F/DSN), OS-Version: 13
    \item Samsung Galaxy S20 (SM-G781B/DS),  OS-Version: 13
    \item Nokia 6.1 (TA-1043 ),  OS-Version: 10
    \item Xiaomi 14T Pro (2407FPN8EG) , OS-Version: 15
    \item  Oppo Reno 12  (CPH2625), OS-Version: 15
    \item Huawei P30 (VOG-L29),  OS-Version: 12
\end{itemize}

\subsection{Results}
The experiments were carried out in a professional digital forensic laboratory equipped with state-of-the-art tools and instrumentation.\footnote{Due to a non-disclosure agreement (NDA), the specific names and versions of the tools and devices used for imaging cannot be disclosed.} All procedures followed established best practices to ensure conditions closely resembled real-world forensic scenarios. For instance, all mobile devices were initially placed in flight mode, and all external communication interfaces, including Wi-Fi and Bluetooth, were disabled before acquisition.

\subsubsection{Acquisition (Parent)} 
Given that most phones and operating systems are somewhat older, no problems were encountered when acquiring the images, i.e., complete Full File System images were successfully obtained on all used devices. As pointed out earlier, the acquisition of S2 (known PIN) is required to enable Developer Mode and turn on USB debugging. 

In two cases, AFU did not work, and we obtained an error. As mentioned in the methodology, in these two instances (Samsung Galaxy A51 (SM-A515F/DSN), Apple iPhone 11 64 GB  black (A2221)), we restarted the phone and brute-forced the PIN / used BFU. 

\subsubsection{Acquisition (Child)} 
For S1 (unknown PIN), the results were identical to the previous experiment, i.e., we obtained Full File System images for all phones. However, for S2 (known PIN), the situation is slightly different.
This type of acquisition requires a change of settings (ADB, USB debugging).
\begin{description}
    \item[iPhone:] No difficulties were encountered for iPhones, i.e., enabling Developer Mode on the iPhones was feasible despite the presence of child restrictions. 
    
    \item[Android:] However, for the Android phones, the parental control mechanisms prohibited enabling the Android Developer Mode. 
Consequently, none of the S2-Android devices could be imaged and highlighting the first impact of the parent controls for practitioners.
The rationale for this restriction is obvious: the child should not possess administrative privileges on the device, and enabling Developer Mode constitutes an action requiring elevated privileges. 
\end{description}

\subsubsection{Data comparison: Parent vs.~child image}
Parent and child images were compared on a per-device basis. Multiple images for the same device were consolidated for analysis. Given the small dataset, manual analysis using a checklist was preferred over automated scripts, with some automated functions employed for hash verification and filtering by tags.

The analysis focused on file integrity, metadata, browser histories, app installation and deletion (e.g., WeChat), deleted content, contacts, calls, calendar entries, emails, and health app data. Minor discrepancies were observed due to differences in operating system builds, imaging software versions, or account restrictions, such as the child account lacking administrative privileges.

On Apple devices, the installation of applications such as Telegram and Discord was not possible under a child account. This restriction originates from Apple's App Store policies rather than the imaging process itself. From a forensic standpoint, it is important to recognize that these applications cannot be installed or used in the standard manner on a child's Apple device.

For Samsung Android devices, a similar pattern was observed. When configured under a child account, certain Samsung applications, such as Samsung Voice, Samsung Notes, and Samsung Health, were not installed during initial setup. In one case, the Samsung Internet Browser was also missing. Again, these absences are the result of account restrictions, not imaging errors. Since children cannot create or link a Samsung Account, services dependent on that account, including Samsung Health, are unavailable in a child environment.

In summary, no substantial forensic differences were identified between the parent and child devices. This conclusion applies both to device pairs linked under the same family configuration and to cross-device comparisons. These findings are consistent with expectations: forensic images remain structurally identical regardless of account type. Any observed variations stem not from imaging discrepancies but from functional restrictions imposed on child accounts, e.g., the inability to install specific applications or access services tied to an account type.

\subsection{Discussion}
With respect to the differences between the parent and child devices, the most relevant factor identified was the inability to activate Developer Mode on the child device. This means, if AFU or BFU are not an option and ADB is deactivated, the data on the device cannot be copied. This raises the question of what options exist.

\begin{figure}[ht]
    \centering
    \includegraphics[width=0.5\linewidth]{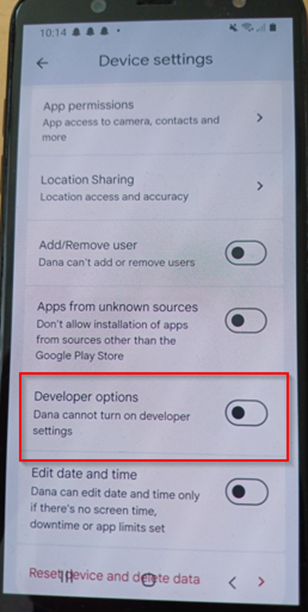}
    \caption{Option within Google Family Link on the parent's device to enable Developer Options on the child's device.}
    \label{fig:placeholder}
\end{figure}

If the parent device is available, the Family Link application can be used to grant permission for the child's phone to enable Developer Mode (see Fig.~\ref{fig:placeholder}). Once this permission has been granted through the parent’s device, the child can activate Developer Mode directly on their own phone. This procedure was tested on the `child' Samsung Galaxy A51 (SM-A515F/DSN).

However, transmitting this setting change to the child device requires connectivity. Thus, the Samsung Galaxy A51 must deactivate flight mode and be connected to Wi-Fi or a mobile network.
From a forensic perspective, this introduces risks: remote deletion may occur, and incoming messages or updates can alter the device’s original state. While feasible, this method is not optimal.

\cite{reiber2019} demonstrated connecting the phone to a computer via a USB cable and copying files directly from the device. While this approach is inelegant from a forensic standpoint, it may be considered when no other options are available. To improve the forensic soundness of such a procedure, the USB interface should be placed in a read-only state where possible (for example, by employing a hardware write-blocker or an OS-level read-only mount). After file acquisition, the examiner should create an archival container (e.g., a ZIP file) and compute cryptographic hash values for the archive and for the individual files to preserve and demonstrate integrity. 
To evaluate the practical applicability of manual extraction on a child's handset, we tested this manual USB method on a Xiaomi 14T Pro. We were able to change the device's USB mode from `charging' to `file transfer' and establish a successful connection to a Windows 11 host system, permitting direct folder access.

A further manual extraction strategy is to capture high-resolution images of the device screen. Forensic camera kits and documented photographic workflows facilitate the capture, storage, and processing of screen images in a manner that supports evidential integrity and continuity of custody.
Other wireless acquisition alternatives include Bluetooth, infrared, NFC, and WLAN-based techniques. These methods require activation of the corresponding interfaces and carry several practical and evidential risks: they may trigger device processes that modify or delete data, permit incoming communications that overwrite volatile information, are relatively slow, and are limited in the scope of retrievable artifacts compared with full logical or physical acquisitions.

\paragraph{Recommendation} The optimal approach involves configuring these settings on the parent's device before activating flight mode. In a forensic context, the most appropriate moment to modify these settings is when the phone is removed from the owner's possession. At that time, the device should not be in flight mode. The recommended procedure is to have the phone examined by an investigator, instruct the parent to enable `Developer options' via the Google Family Link application on their device, then access the `Developer options' on the child's phone. Subsequently, the device should be set to flight mode, initiating the chain of custody with comprehensive documentation of all actions taken before this point. 
If this procedure can be completed on an Android device before it departs from the scene of the incident, it represents the most effective preparation for ensuring the forensic soundness of the child's phone and maximizing its accessibility for subsequent analysis. Additionally, to ensure all information is accurately preserved, it is advisable to request the parents' parental control PIN code. 
If the parents are not available or do not cooperate, it is advisable to perform AFU or BFU imaging as a second possible option.

\subsection{Additional considerations for extracting data} 
Depending on the specific requirements of an investigation, pertinent information can potentially be obtained from the parental control application. The application provides details regarding the child's device, such as visited locations and the periods during which the device was actively used. If the inquiry pertains to a particular time frame or location, this could serve as an alternative means of gathering relevant information. 

Alternative invasive methods for data collection include JTAG or elevating root privileges. Since Android is based on a Linux kernel, obtaining root access is feasible \citep{reiber2019}. Many forensic tools that deploy an APK onto the target device rely on shell root privileges to enable the creation of a forensic image. Shell root access is temporary, i.e., after a system reboot, elevated privileges are revoked and the device reverts to its standard configuration. In contrast, permanent root access persists across reboots, thereby increasing the device’s exposure to potential malware infections and compromising its overall security posture \citep{reiber2019}. 

Discussions in various online forums indicate the existence of additional, often ambiguous techniques that require advanced hacking skills or the use of unverified tools. These approaches diverge significantly from forensically sound methodologies and therefore fall outside the scope of this study.

All three parental control applications reveal that users have identified various `tricks' to circumvent the parental control functions. The most prevalent methods include altering the birth date \citep{heise2025, wfaulconer2024}, storing data in secure folders, or activating accessibility features intended for individuals with disabilities \citep{schanze2024, heise2024}. For Windows devices, in addition to modifying the birth date, users often employ techniques such as creating additional accounts or changing the system time \citep{dasr2024}. These steps do not effectively address the core issue of activating Developer Mode.


\section{Experimental design and results for the Windows devices}
\label{sec:experiment_windows}
Like the previous section, the aim is to assess the impact of parental control on digital forensics. While the previous section focused on phones, this one assesses Windows devices. Overall, we followed a similar approach. Next we will discuss the methodology (Sec.~\ref{sec:metho_win}) followed by \nameref{sec:win_devices_conf} and the results (Sec.~\ref{sec:results_windows}).

\subsection{Methodology}\label{sec:metho_win}
The procedure for the device was similar to the phones.

\begin{enumerate}
    \item \textbf{Initial Configuration (Parent):} The tested Windows device was configured according to our playbook summarized in Sec.~\ref{sec:win_devices_conf}, and left in parent mode.

    \item \textbf{Acquisition (Parent):} We acquired the data from the Windows device, i.e., triggering the imaging process. We simulated two situations (S):
    \begin{enumerate}
        \item S1: System running 
        \item S2: System turned-off
    \end{enumerate}

    \item \textbf{Reset:} We formatted the Windows device to start with a clean device
    
    \item \textbf{Initial Configuration (Child):} Like in the initial configuration (Parent) phase, we reconfigured the Windows device, but this time in child mode.
    
    \item \textbf{Acquisition (Child):} Similar to step 2, we attempted to copy the data under the identical conditions, i.e., S1 and S2. Comparing this step to step 2 allows us to compare the acquisition procedure.

    \item \textbf{Data comparison:} Lastly, we compared the images that were acquired, i.e., the image parent vs.~image child, to see if there were any differences in data. 

\end{enumerate}

\subsection{Devices and their configuration} 
\label{sec:win_devices_conf}

Each device was configured with a standardized setup and a documented device protocol (see Appendix). 
Devices were populated with frequently examined data types, including pictures, videos, audio files, and documents, all loaded via USB and sent by email. Browsers (Chrome, Firefox, TOR, with Microsoft Edge pre-installed) were installed and used to access identical websites, including YouTube, Google Maps, and Minecraft.net; TOR was later uninstalled. Windows applications such as Outlook, Calendar, and Notes were also used to generate content. During media import and imaging, two USB sticks were connected, and selected pictures and videos were deliberately deleted to simulate typical usage.
We used four identical Dell OptiPlex 7070 devices running Windows 11 (Build 24H2). Using multiple devices allowed us to compare results and confirm that observed behaviors were consistent, rather than anomalies from a single device reacting differently.

\subsection{Results}
\label{sec:results_windows}
As with the telephones, the study was conducted in a digital forensic laboratory using state-of-the-art tools and following best practices.

\subsubsection{Acquisition (Parent)}
The parents' devices were initially imaged in a live forensic acquisition scenario, wherein a USB drive was inserted into the target computer, and a physical image was created using FTK Imager (S1). 
In the subsequent phase, the system was powered down (S2), and the solid-state drive (SSD) was extracted and imaged directly using the Tableau Forensic Imager to obtain a physical image.
In both scenarios, the imaging process was completed successfully without any errors. The resulting image files were verified through hash value validation to ensure data integrity. 
For both acquisition variants, the BitLocker encryption key was present on the devices and was subsequently required to access and process the imaged data within Autopsy.

\subsubsection{Acquisition (Child)}
The acquisition process was first attempted while the system was running (S1). This was followed by a second acquisition performed after powering down the device and imaging the extracted SSD (S2).

The \emph{live forensic acquisition (S1)} is not possible using FTK Imager. This limitation stems from the device configuration, wherein the Windows child-user account is restricted to standard user privileges without administrative rights. 

With respect to the \emph{physical image extraction process (S2)}, no distinction was observed between the child's device and the parent's device; both procedures were technically feasible and successfully executed. However, an obstacle arose following the imaging phase when attempting to retrieve the BitLocker recovery key. Accessing this key requires either the parental control PIN set by the parent or the parent's account credentials. This security mechanism parallels the principles underlying Developer Mode in Android systems: a child account is designed to operate without administrative privileges and therefore functions as a standard, non-administrator user within the Windows environment.

\subsubsection{Data comparison: Parent vs.~child image}
The Windows devices were analyzed using Autopsy with all available Ingress modules. As with the mobile device analysis, parent and child images on the same device were compared.

Analysis was performed manually using a checklist 
and included verification of all previously established settings and optional elements, such as images with hash-sets. The investigation focused on several key questions: presence and integrity of media files, differences in metadata, visibility of deleted files, browser usage and search histories, USB device insertions, email accounts, calendar entries, and sticky notes.

Overall, parent and child images were largely identical. Minor differences were observed due to the child account lacking administrative privileges, affecting the visibility of certain elements. For example, USB devices were fully visible in parent (administrator) accounts but not in the child account. However, these devices could still be identified via system files (e.g., usbstor.inf). This pattern was consistent across all cases. Apart from such account-based differences, no significant discrepancies were noted.

\subsection{Discussion}
Regarding Windows devices, we have identified two crucial peculiarities in the imaging process. 
First, a powered-on device cannot be imaged using a forensic tool that requires administrator rights. While we initially tested FTK Imager, in a subsequent test, we also tried other forensic tools such as Autopsy, Kape Kroll, Magnet RAM Capture, and Sysinternals Suite from Microsoft, but all required administrative privileges to initiate imaging. Various forensic environments, such as Tsurugi Acquire, SIFT, and CAINE, were observed, but these environments require a reboot of the device, which destroys the RAM content of the running device, i.e., the live forensic aspect. 
Second, a powered-off device necessitates the BitLocker key. Access to this key requires either a parental PIN or parental account credentials. Consequently, without administrative privileges, it will be difficult to obtain all data/perform a forensic copy. 

One obvious possibility for collecting data involves manually browsing the device and capturing relevant details by taking pictures or videos. As noted with the Android devices, connecting a USB stick and storing selected folders or ZIP files can serve as a contingency to preserve critical information. Any actions that modify the device must be carefully documented to ensure a clear record of all changes.

\paragraph{Recommendation} 
To obtain a forensic image of a Windows device, administrative rights are indispensable. The most straightforward way to acquire these privileges is through the cooperation of the parent, who can provide their Windows PIN. With this PIN, forensic tools such as FTK Imager can be used to create a complete and verifiable image under full administrative access. This procedure was tested on an OptiPlex system and functioned correctly, producing results equivalent to those obtained under a standard administrator account.
The PIN is also required to retrieve the BitLocker encryption key of the device. 

An alternative to using the PIN for initiating the imaging software is to switch to the account that possesses administrative rights (without logging out the child's account) and to perform the imaging process from this account, while the child's account remains active in the background. 

If an administrator PIN is unavailable, only tools that run without elevated privileges may be executed. Parts of the NirSoft suite (Nir Sofer) can be useful in this non-elevated context: for example, BrowserDownloadsView enumerates browser downloads and reports the download URL, source web page URL, and download start/end timestamps; BrowserHistoryView lists visited URLs, the originating browser, and visit timestamps (note: Tor Browser history is not exposed). Both utilities support exporting results in plain-text/CSV formats suitable for case documentation and further analysis.

However, several NirSoft components require administrative rights (e.g., FileActivityWatch.exe) and therefore cannot be used from a restricted account. In addition, Windows Defender commonly flags these utilities as potentially malicious, and their execution may alter the system state. For these reasons, any use must be carefully documented by the forensic examiner to preserve the integrity of the investigation.

\subsection{Additional considerations} 
\label{sec:otheroptions}
Microsoft collects and displays activity reports through the Microsoft Family Safety platform. These reports are limited in scope: they only track activity when the feature is enabled and record actions within Microsoft applications. For example, activity performed in Microsoft Edge, including search terms and visited websites, is visible, whereas activity in other browsers, such as Firefox, is not. Reports are retained for a maximum of two weeks.

If accessible, a parent-submitted activity report can provide useful investigative information, including recent searches, visited or blocked websites, and the most-used applications. For instance, such a report may reveal that a child used the Tor Browser and conducted searches like `python programming' or `Minecraft' via Edge. With parental cooperation, these reports can serve as an alternative source to answer targeted questions, such as whether a particular website was visited or a specific location was searched.

\section{Cloud (data) takeouts with the children's accounts}
\label{sec:cloudtakeouts}
A cloud takeout (a.k.a.~data takeout) refers to the process of exporting user data directly from a cloud service provider, such as Google, Apple, or Microsoft. These exports enable the retrieval of account-linked artifacts without requiring physical access to the device itself (best-case scenario if no 2FA is activated). Cloud takeouts can contain a wide range of significant material, including photographs, videos, emails, contacts, calendar entries, and application artifacts that may not be accessible on the physical device or may have been deleted locally \citep{interpol2021, evidenceproject2013}. 

For instance, \cite{ball2018}'s analysis of Apple’s iCloud takeout demonstrates this evidentiary value. His exported dataset included not only live content but also deleted items, full-resolution images rather than thumbnails, and embedded metadata such as EXIF geolocation information. Ball reported that Apple’s export consisted of roughly 63 GB of data delivered in 26 ZIP archives, assembled within approximately five days. These archives are provided in structured formats such as JSON, CSV, or VCF, which preserve data integrity and facilitate forensic parsing. 
When valid account credentials (username and password) are available, cloud takeouts can serve as an effective alternative or complement to conventional mobile forensic techniques. 

The process of acquiring cloud data, however, is not without complications. As \cite{reiber2019} notes, cloud extractions are technically complex and can involve multiple layers of authentication. Two-factor authentication (2FA) mechanisms, such as PINs, tokens, or push notifications, can obstruct acquisition unless the examiner has access to the target device or associated authentication factors. Some cloud services also generate automatic email or system notifications when a takeout is initiated, which may alert the account holder. In cases where the user is already authenticated on a device, investigators may leverage existing session tokens, though these are temporary and may expire. Available targets for extraction include provider-level services (e.g., Apple iCloud, Google Account data) and application-specific stores (e.g., Snapchat, Fitbit).

The most forensically sound approach to performing a cloud takeout is through forensic tools that automate data acquisition, preserve metadata, and generate verifiable hashes and audit logs. If specialized tools are unavailable, a provider’s native export mechanism (e.g., Google Takeout or Apple’s privacy portal) may be used as an alternative, provided that the procedure is documented and conducted in a controlled environment to preserve integrity. The legal authority to access and process cloud-hosted data varies by jurisdiction and therefore falls outside the scope of this study.

To evaluate whether cloud evidence could serve as a viable alternative or complement to the `image evidence', cloud takeouts were conducted for children's accounts. It is particularly noteworthy to examine whether additional restrictions are imposed when creating a cloud takeout from a child's account. 
Although a manual extraction was performed in this study, several forensic tools are available that support the access and download of such cloud-based data. 

As pointed out earlier, we did not opt for 2FA when setting up the accounts.

\subsection{Google Takeout} 
\label{sec:googletakeout}
The Google Takeout process can be initiated via \url{https://takeout.google.com/?pli=1} and requires the username (email address) and password. The extraction was possible without parental consent or PIN. 

An examination of the provided data reveals various details of the device itself. The takeout data, for instance, displays the device model and type at several points. From a mobile device perspective, numerous details can be inferred, including the IMEI (International Mobile Station Equipment Identity), phone model, and display information such as screen height. 
Furthermore, within the Chrome section, there is information regarding the Chrome version, bookmarks, browser history, and the dictionary. Concerning our testing configuration, the entire browser history is displayed accurately.  

The `Google Photos' folder stored pictures and their associated metadata. Only photographs taken with one's own camera are included. Furthermore, the content varies among the three children's accounts; only one account contains pictures, while the other two folders were empty. We were unable to find out why this inconsistency happened. 

Within the Gmail section, all emails present in the children's original mailbox were displayed. Additionally, all Google contacts and calendar appointments were visible. The `Google Play Store' section listed all installed applications along with their installation dates. It also indicated family approval, i.e., it was visible when a parent had authorized the installation of WeChat. 

The section titled `Google account' displayed a change history and subscriber information. It confirmed that the account is an official Google child account, identified by tags from Google Services for children, including `Is Unicorn Child Account', `Is Restricted Account', `Is Supervised Account', and `In Family'. Furthermore, the section `Access log activities' presented activity logs for applications such as Chrome and others.

Elements outside of Google applications, such as Discord contacts, WhatsApp pictures, and others, are not visible. Similarly, notes and audio files are also unavailable. 

In the absence of two-factor authentication and with access to the email and password of the Google account, Google Takeout may serve as a valuable alternative to a traditional forensic image, particularly when only an `image with a PIN' is feasible and Developer Mode cannot be activated. Its suitability depends on the information an investigator seeks. When content from third-party applications is relevant, Google Takeout is not an option. Conversely, if contacts, emails, or browsing history are of interest, Google Takeout is a viable solution.

\subsection{Apple Takeout} 
\label{sec:appletakeout}
The extraction request is initiated through the Data and Privacy page accessible at \url{https://privacy.apple.com/}. There, the child is prompted to log in using their username and password, followed by two-factor authentication on the iPhone. The Apple Takeout service is only available for a child who has parental consent, provided via the parent's device. Subsequently, a data request could be made. 

The data provided displays device information such as device name, IMEI, serial number, last heartbeat IP address, ICCID (Integrated Circuit Card Identification Number), and IMEI. Additionally, it included sign-in information such as login dates, IP addresses, and details of installed applications. Bookmarks and favorites from the Safari browser were also shown, along with all calendar entries in file format and all contacts stored within the native phone contacts. Notes were present with their content, indicating whether they were pinned, modified, or deleted.  

The `iCloud Drive' located in our child test case displayed all available test files, such as pictures, videos, and audio recordings from the iPhone. This included files created directly on the device as well as downloaded content. The deleted file was present in the trash folder (it likely would be deleted after a certain time, e.g., 30 days). In addition to the files themselves, the document provided details including title, base hash, file type, size, creation date, modification date, and last opened date. Within the folder `iCloud Photos – Albums', an Excel spreadsheet named `WhatsApp' lists the titles of three JPG files.
Besides images from WhatsApp and the list of their associated names, no content related to third-party applications is visible, i.e., no chats, images, or audio from applications such as Instagram or TikTok were accessible. 

Considering the content, if the metadata from photos, calendar entries, notes, or emails is of interest, the takeout could be relevant to the investigators. However, Apple Takeout does not represent the most forensically sound option as it requires disabling flight mode on the child's iPhone to facilitate two-factor authentication (in addition, parental consent on the guardian's device is needed). The acquisition of a forensic image should, therefore, take precedence.

\subsection{Windows Takeout} 
\label{sec:windowstakeout}
A Windows Takeout is initiated from the following link: \url{https://www.microsoft.com/en-us/privacy/privacy-support-requests}. There, the option `Make a request related to a personal Microsoft account' is selected. This selection redirects the user to their own account page, where the `Submit a Privacy Request' section is accessible. 
Login was possible with the email and password of the child's account.  

Generally, the Windows Takeout contained less information than the Google/Apple Takeouts. We were able to view the Edge browser history (no other browser), which included the time, page title, domain, full URL, and application data, but not much other information, i.e., its capabilities were quite restricted for tasks such as identifying an IP address or determining whether a user was online. Images or videos were not included in the takeout but may be obtainable with direct access to OneDrive. Due to the limited scope of the test cases, several folders and files remained empty but may contain data when being used over a longer period of time. Additional testing is needed.

In summary, the Windows Takeout provided little information and should be considered merely as one of several options, to be used only when no other methods are feasible.



\section{Limitations}\label{sec:limitations}
This study was conducted under controlled laboratory conditions, which do not fully replicate the diversity and unpredictability of real-world forensic environments. The experiments were limited to a predefined set of devices and configurations, i.e., specifically selected Android, iOS, and Windows systems, and therefore may not reflect differences introduced by other device models, operating system versions, or regional configurations.
All experiments were carried out in Switzerland, and it should be noted that the functionality and scope of parental control systems may vary across jurisdictions due to differing regional regulations and service implementations.
Furthermore, the data comparison and validation processes were conducted manually, following standardized written procedures. While every effort was made to ensure accuracy and consistency, the potential for human error cannot be entirely excluded.

\section{Key Insights}
\label{sec:Key_Insights}
This section revisits the research questions raised in the introduction and then summarizes the key takeaways.

\subsection{Answers to the research questions}
While the questions were discussed in detail throughout the experiments, the following provides a summary per RQ.

\textbf{RQ1: To what extent do parental controls hinder the work of digital forensics?}
We evaluated iOS and Android test devices to assess whether parental controls hinder forensic acquisition in either the AFU or BFU states. None of the AFU and BFU processes were influenced by the parental controls. However, it impacts `known PIN' acquisitions via ADB if the Developer Mode is not activated. It also impacts the acquisition of Windows devices.

\textbf{RQ2: To what extent do parental controls affect the quality of forensic images?} 
Our experiments show that parental controls have only a minor effect on the quality of forensic images. The results of the data analysis show that there were minor discrepancies in the data; we estimate that this will have only a minimal impact on the results of an investigation.

\textbf{RQ3: How can parental controls be disabled or bypassed in the most forensically sound manner?}
We proposed ideas for bypassing parental controls that can be regarded as forensically defensible. However, challenges have been identified in bypassing Android Developer Mode and in obtaining administrative rights on Windows 11. A truly forensically defensible method for bypassing such protections cannot be achieved.

\subsection{Considerations and key takeaways}
This section provides some additional thoughts and key takeaways from our study, considering the peculiarities of the different operating systems.

\subsubsection{Android} 
For Android devices, both AFU and BFU imaging are feasible acquisition methods and should be considered among the most effective strategies for mitigating potential complications posed by parental controls during forensic imaging. Even when the device PIN is known, parental control settings typically restrict the activation of Android Developer Mode. Therefore, AFU and BFU imaging are advised as primary acquisition strategies; when appropriate configuration parameters for PIN iteration are available, a controlled brute‑force procedure may be considered as an additional option. 
The most appropriate approach to resolving the issue of enabling Developer Mode (while maintaining adherence to defensible forensic principles) is for a digital forensics practitioner to convince the parent or legal guardian to authorize the activation of Developer Mode through the parental control application, before placing the device in flight mode. Furthermore, the extraction of cloud data via Google Takeout may constitute a viable alternative for resolving unresolved investigative inquiries, particularly when access to the device is limited due to active parental control settings. The accounts examined were able to generate a Google Takeout using only the associated email address and password (2FA was not activated). With regard to this method, only fundamental Google-related data, such as Chrome browser history and account login activity, can be retrieved, while information associated with third-party applications remains inaccessible. If other approaches are infeasible, manual extraction of artifacts such as chat messages and images may be undertaken. 

\subsubsection{Apple iPhones} 
On Apple iPhones, the imaging process is not adversely affected by parental controls. Required configurations, such as enabling Developer Mode, can be performed without impediment. However, certain applications, including Telegram and Discord, were neither installable nor accessible on the tested devices due to parental control restrictions. Alternative acquisition methods, such as Apple Cloud Extraction, cannot be reliably recommended, given that two-factor authentication and parental control agreements are prerequisites for this process. It is therefore advisable to verify that the parental control application’s PIN is operational and accessible to ensure the full functionality of all relevant features during forensic examination.

\subsubsection{Windows 11 devices} 
On Windows 11 devices, the creation of a powered-down image via direct imaging of the hard disk or SSD is not directly affected by parental controls. Access to the BitLocker recovery key is restricted to an administrator account, i.e., in this context, the parent account—necessitating the corresponding PIN or password to ensure its retrieval. In live forensic scenarios, tools requiring administrative privileges will be rendered ineffective due to the absence of administrator rights on the child's user account. This can influence the location and the way data is stored. If a Windows 11 device is powered on, an image without administrative rights cannot be created without the PIN of the parent or administrator account. 
Consequently, it is strongly recommended to obtain the parental control PIN from the custodial parent or guardian. Additionally, alternative investigative avenues such as Microsoft Cloud Takeout or examination of parental control activity reports may provide supplementary insights, particularly with respect to browser history. However, visibility is limited to Microsoft-native applications, such as the Edge browser, while third-party applications, including Firefox, remain inaccessible within these reports.

\section{Conclusion}
\label{sec:conclusion}
This study explored the impact of parental control technologies on mobile devices (Android, iOS) and Windows systems in the context of forensic investigations, while proposing approaches to mitigate these challenges. The key findings indicate that parental controls can hinder standard forensic procedures, e.g., a running Windows device in child mode does not have administrative privileges and cannot be copied; Android devices typically prevent activation of Developer Mode. To our knowledge, this is the first study to examine how parental control systems affect investigative processes, despite their widespread use over recent years.

The study has limitations, including a small sample of 15 devices, controlled experimental conditions that do not fully replicate real-world usage, and a focus on a single country, which may limit generalizability. Future work should expand device samples, incorporate more realistic user interactions, and consider cross-jurisdictional variations to better understand and address the challenges posed by parental control technologies in digital forensics.



\section*{Disclosure of AI-assisted writing tools} 
Some authors used Grammarly and ChatGPT to assist in revising, condensing text, and correcting grammatical errors, typos, and awkward phrasing. All AI-generated suggestions were carefully reviewed and modified as necessary to ensure they aligned with the authors' intended meaning before being incorporated into this article.

\section*{Declaration of interest}
The authors declare that they have no known competing financial interests or personal relationships that could have appeared to influence the work reported in this paper.

\bibliographystyle{model5-names}
\bibliography{refs}

\end{document}